\documentclass{iaus}
\usepackage{graphicx}

\title[Be/Oe stars and LGRBs progenitors] 
{ZAMS rotational velocities of Be/Oe stars and LGRBs progenitors in the Magellanic Clouds.}

\author[C. Martayan \& J. Zorec \& Y. Fr\'emat]   
{Christophe Martayan$^{1,2}$
 \and Juan Zorec$^3$
 \and Yves Fr\'emat$^{1,2}$
 }

\affiliation{$^1$Royal Observatory of Belgium, 3 avenue circulaire \\ 1180 Brussels, Belgium 
 \\ email: {\tt martayan@oma.be} \\[\affilskip]
$^2$GEPI, Observatoire de Paris, CNRS, Universit\'e Paris Diderot; 
5 place Jules Janssen \\ 92195 Meudon Cedex, France \\[\affilskip]
$^3$Institut d'Astrophysique de Paris, UMR7095, CNRS, Universit\'e Marie \& Pierre Curie, \\ 98bis Boulevard Arago 
75014 Paris, France \\[\affilskip]
}

\pubyear{2008}
\volume{256}  
\pagerange{119--126}
\jname{The Magellanic System: Stars, Gas, and Galaxies}
\editors{Jacco Th. van Loon \& Joana M. Oliveira, eds.}
\begin{document}

\maketitle

\begin{abstract}
The Large and Small Magellanic Clouds are priviledged environments to
perform tests of theoretical predictions at low metallicity on rotational
velocities and stellar evolution. According to theoretical predictions, the rotational velocities of
B-type stars are expected to be higher in low metallicity (LMC/SMC) than in high metallicity (MW) environments.
To verify the models, we observed with the VLT-FLAMES 523 B and Be stars,
which form, at the moment, the largest observed sample of these kind of objects in the MCs.
We first determined the stellar fundamental parameters and we found that B and Be stars rotate 
faster in the MCs than in the MW.
We also determined the first distribution of the average ZAMS rotational velocities versus the mass of Be stars. 
These results indicate that the appearance of Be stars is mass-, metallicity-, stellar evolution-, 
and star-formation regions-dependant.
Moreover, the recent models of Long Gamma Ray Bursts progenitors foresee possible LGRBs progenitors 
at the SMC's metallicity. We confront these models with the observed (ZAMS rotational velocities, masses) 
distributions of the fastest rotators (Be and Oe stars) in our sample. Furthermore, we compare the corresponding
predicted rates from our study with observed rates of LGRBs. 

\keywords{stars: emission-line, Be, Magellanic Clouds, stars: fundamental parameters, gamma rays: theory}
\end{abstract}

\firstsection 
\section{Introduction, observations}

This document deals with the determination of rotational velocities of B-type stars at different metallicities (MW, LMC,
SMC) and their consequences on the stellar evolution in particular for the late stages. A comparison between the results
from the observations and theory is also given.
To perform this study, thanks to the VLT-FLAMES/GIRAFFE facilities in MEDUSA mode, 
a large sample of B and Be stars have been observed in the LMC-NGC2004 and SMC-NGC330 regions and their surroundings, 
176 and 344 hot stars respectively. We used a medium resolution R= 6400 
in the blue wavelength range around H$\epsilon$, H$\delta$, H$\gamma$ (LR02) and R=8600
in the red one around H$\alpha$ (LR06). Among this sample, 178 Be stars were observed.

\section{Rotational velocities vs. metallicity: observations and theory}
 
Using the GIRFIT code (Fr\'emat et al. 2006) we determined the fundamental parameters for each star by
fitting observed spectra with theoretical NLTE spectra. The gravitational darkening effects 
for fast rotating stars
such as Be stars were taken into account thanks to the FASTROT code (Fr\'emat et al. 2005).  

Statistically, Martayan et al. (2006, 2007) found that B and Be
stars with similar masses and ages rotate faster at low metallicity (in the SMC) than at high metallicity
(in the MW). This result was the first confirmation, based on a significantly large star sample, of the
theory from Maeder \& Meynet (2001) about the behaviour of rotational velocities in respect of the
metallicity for main sequence (SMC vs. MW) B and Be stars. Hunter et al. (2008) recently arrived at the
same conclusions. Our results in the LMC are in agreement with those of Keller (2004).

\begin{figure}[h!]
\begin{center}
 \includegraphics[width=2.7in, angle=-90]{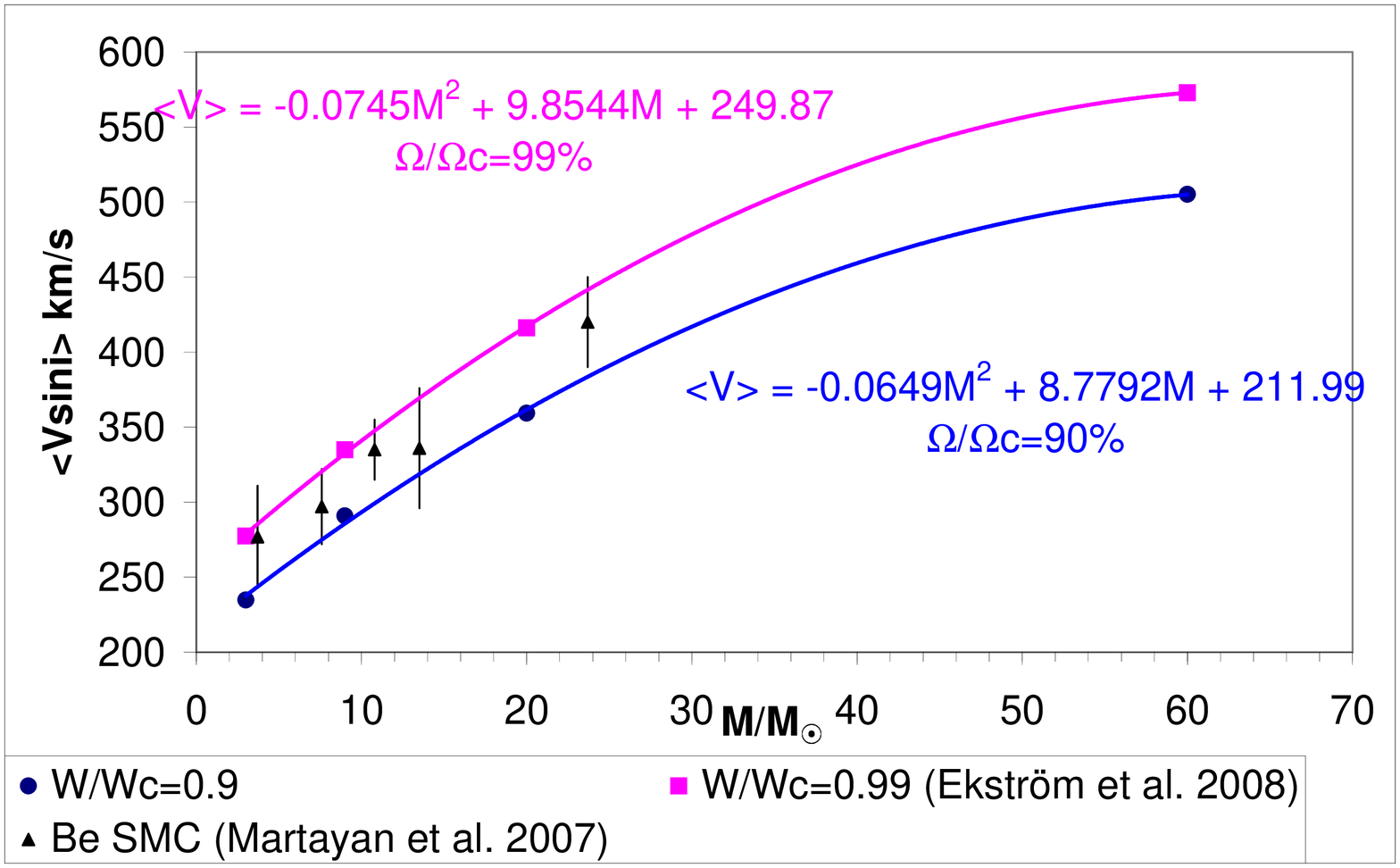} 
 \caption{Comparison of average Vsini distribution for observed Be/Oe stars in the SMC (black points from Martayan et al.
 2007) with the theoretical tracks for stars, which display the Be-phenomenon at the SMC metallicity (pink and blue curves
 are from Ekstr\"om et al. 2008).}
   \label{fig1}
\end{center}
\end{figure}

More recently, Ekstr\"om et al. (2008) published new stellar evolutionary tracks at the SMC metallicity. 
In Fig.~\ref{fig1}, we compare the independent results from Martayan et al. (2007), for which we add the values for the most
massive SMC Be stars-late Oe stars, with the results from Ekstr\"om et al. (2008) for the Vsini
distributions. As shown, there is a good agreement between the two studies and the theory seems to correctly reproduce
the observations.

\section{ZAMS rotational velocities}
 
Using the observed values from Martayan et al. (2007) and evolutionary tracks from Schaller et al. (1992,
SMC and MW metallicities) and Charbonnel et al. (1993, LMC metallicity), Martayan et al. (2007) also determined 
for the first time the ZAMS rotational velocities of Be stars in the SMC, LMC, and MW. The
values for the linear rotational velocities in the MW come from Zorec et al. (2005).

\begin{figure}[h!]
\begin{center}
 \includegraphics[width=2.7in, angle=-90]{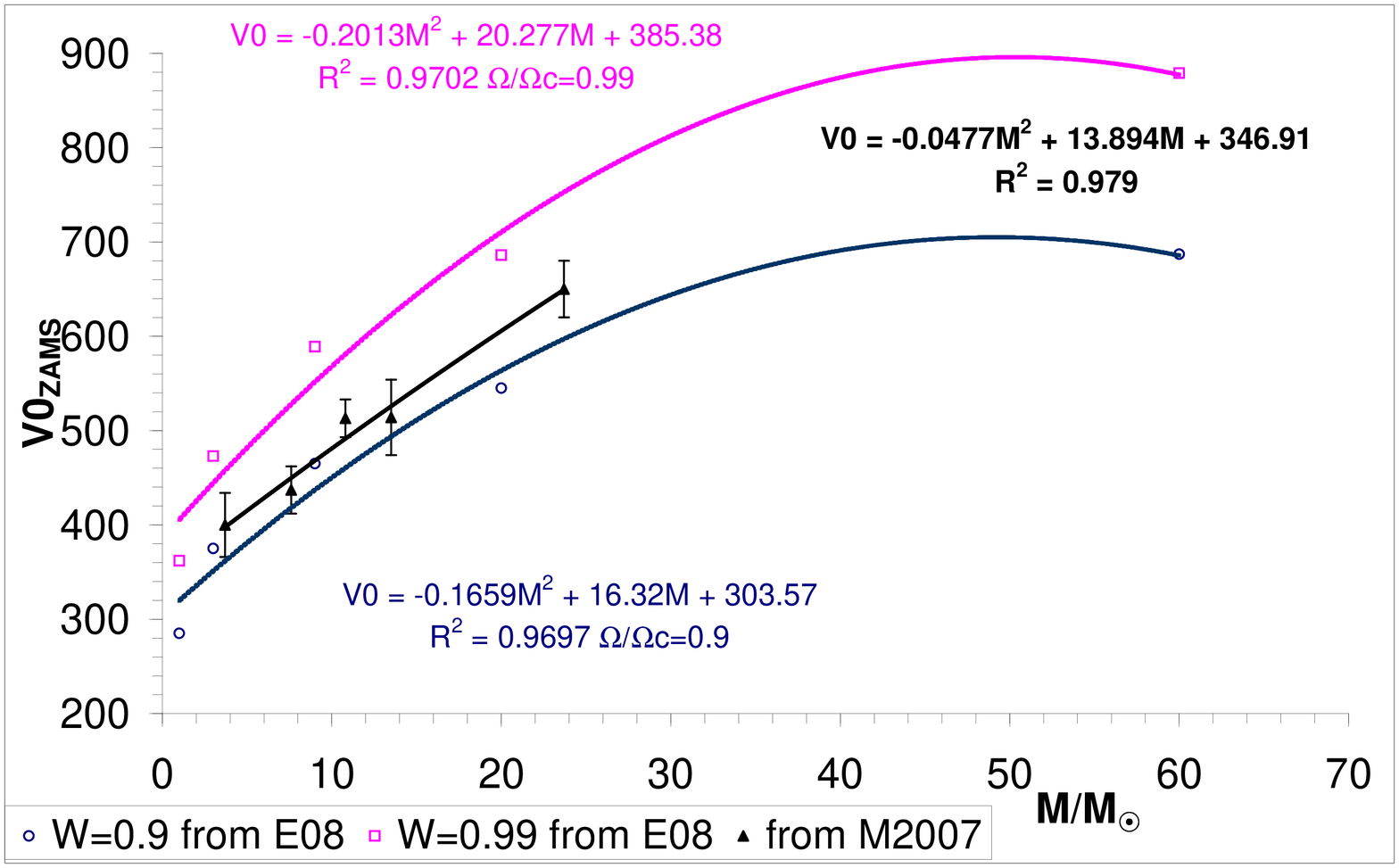} 
 \caption{ZAMS rotational velocities distributions for observed Be/Oe stars in the SMC (black points and curve) 
 compared with theoretical distributions from Ekstr\"om's et al. (2008). }
   \label{fig2}
\end{center}
\end{figure}

In Fig.~\ref{fig2}, we add the ZAMS rotational velocities for massive SMC Be and late Oe stars to the values 
from Martayan et al. (2007). We fitted 2nd order polynomials through these values. 
We compared our observational values to
Ekstr\"om's et al. (2008) theoretical ones. The comparison shows again a good agreement between the theory
and observations. The distributions show an increase of V0$_{ZAMS}$ for B stars and late O stars (due to low
mass-loss) and a decrease of V0$_{ZAMS}$ at the more massive O stars (in which the stellar winds are strong
enough to lead strong mass-loss and strong angular momentum-loss). However, we need observations of more
massive stars, like O fast rotators stars to better constrain this trend.  

\section{Stellar evolution and Long Gamma Ray Bursts progenitors}
 
The Gamma Ray Bursts are the most energetic events in the universe since the Big Bang. We distinguish two
kinds of GRBs, the Short and Long soft GRBs. It is believed that Short LGRBs can be produced by the merging
of two compact objects such as neutrons stars in a binary system. 

Here we focus on the Long soft GRBs. Since the model of the collapsar from Woosley (1993), where
LGRBs are produced by the collapse of massive fast rotating stars, the recent developments from
Hirschi et al. (2005) or Yoon et al. (2006) predict the ranges of masses and rotational velocities of
the LGRBs progenitors as a function of the metallicity. 
The distributions of ZAMS angular rotational velocities of Be stars in our SMC sample as a function of mass 
with the predicted area of LGRBs by Yoon et al. (2006) show that 
most of the massive Be stars and late Oe stars are located in the LGRBs progenitors' area.

However, Hirschi et al. (2005) foresee more LGRB progenitors from massive fast rotators stars than Yoon et
al. (2006). As a consequence of these studies, we propose that the most probable LGRB progenitors are the
low metallicity (Z of SMC or lower) early Be stars and late Oe stars. Moreover, after the main
sequence, these stars must become fast rotating WR stars. To do that, it is necessary that the stars at the
end of the main sequence be fast rotators and massive stars. The finding by Martayan et al. (2007) that 
massive Be stars at low metallicity are in the second part of the MS is an argument that favours this
scenario.

\section{Be/Oe stars as Long Gamma Ray Bursts progenitors: estimated rates}
 
To test our scenario that takes massive Be stars and late Oe stars as progenitors of LGRBS, we determined
the corresponding rates of LGRBs it implies. To do that, we use the proportions of SMC Be/Oe stars to B/O
stars from Martayan et al. (2008, in preparation).  We also use the number
of B and O stars in the SMC estimated from OGLE catalogues for the whole SMC. We take into account the
lifetime of the stars and the angle of $\gamma$-ray beams to infer the rates of LGRBs events. We obtain that they
range as:  3. 10$^{-6}$ to 3. 10$^{-5}$ LGRB/year/galaxy.

These rates must be compared with observed rates of LGRBs:  from Hirschi et al. (2005, and references
therein): 3. 10$^{-6}$ to 6. 10$^{-4}$ LGRB/year/galaxy, and from Fryer et al. (2007, and references therein) in the
local universe (z$<$1), the values range from: 10$^{-6}$ to 10$^{-5}$ LGRB/year/galaxy. 

The rates we estimated are in good agreement with the observed rates of LGRBs and as a consequence, massive
Be stars and late Oe stars are then among the most probable candidates of the LGRBs progenitors.

\section{Conclusions}
 
The comparisons of rotational velocities distributions of Be stars from our VLT-GIRAFFE observations with
theory show good agreements. The results are also similar for the ZAMS rotational velocities, while we
added the most massive Be stars to the distributions. From the comparison of stellar evolution models to
our observations of SMC Be stars we suggest that the most probable candidate of LGRB progenitors can be the
more massive Be stars and late Oe stars. The rates of LGRBs we estimate with the distributions of Be stars
at low metallicity are in good agreement with the observed rates of LGRBs in the local universe.

\begin{acknowledgements}
C.M. acknowledges funding from the ESA/Belgian Federal Science Policy in the 
framework of the PRODEX program (C90290).
C.M. thanks the IAUS SOC/LOC for the IAUS grant and  the FNRS for the travel grant.
\end{acknowledgements}


\begin{thebibliography}{}

\bibitem[Charbonnel et al. (1993)]{}
{Charbonnel, C., Meynet, G., Maeder, A., et al.} 1993, 
\textit{A\&AS} 101, 415

\bibitem[Ekstr\"om et al. (2008)]{}
{Ekstr\"om, S., Meynet, G., Maeder, A., et al.} 2008, 
\textit{A\&A} 478, 467

\bibitem[Fr\'emat et al. 2005]{}
{Fr\'emat, Y., Zorec, J., Hubert, A.-M., et al.} 2005, 
\textit{A\&A} 440, 305

\bibitem[Fr\'emat et al. 2006]{}
{Fr\'emat, Y., Neiner, C., Hubert, A.-M., et al.} 2006, 
\textit{A\&A} 451, 1053

\bibitem[Fryer et al. (2007)]{}
{Fryer, C.~L., Mazzali, P.~A., Prochaska, J. et al.} 2007, 
\textit{PASP} 119, 1211

\bibitem[Hirschi et al. (2005)]{}
{Hirschi, R., Meynet, G., Maeder, A.} 2005, 
\textit{A\&A} 443, 581

\bibitem[Hunter et al. (2008)]{}
{Hunter, I., Lennon, D.~J., Dufton, P.~L, et al.} 2008, 
\textit{A\&A} 479, 541

\bibitem[Keller (2004)]{}
{Keller, S.~C.} 2004, 
\textit{PASA} 21, 310

\bibitem[Maeder \& Meynet (2001)]{}
{Maeder, A. \& Meynet, G.} 2001, 
\textit{A\&A} 373, 555

\bibitem[Martayan et al. (2006)]{}
{Martayan, C., Fr\'emat, Y., Hubert, A.-M., et al.} 2006, 
\textit{A\&A} 452, 273

\bibitem[Martayan et al. (2007)]{}
{Martayan, C., Fr\'emat, Y., Hubert, A.-M., et al.} 2007, 
\textit{A\&A} 462, 683

\bibitem[Martayan et al. (2008)]{}
{Martayan, C., Baade, D., Fabregat, J.} 2008, 
\textit{IAUS256 proceedings} this issue

\bibitem[Schaller et al. (1992)]{}
{Schaller, G., Schaerer, D., Meynet, G., et al.} 1992, 
\textit{A\&AS} 96, 269

\bibitem[Woosley (1993)]{}
{Woosley, S.~E.} 1993, 
\textit{ApJ} 405, 273

\bibitem[Yoon et al. (2006)]{}
{Yoon, S.-C., Langer, N., Norman, C.} 2006, 
\textit{A\&A} 460, 199

\bibitem[Zorec et al. (2005)]{}
{Zorec, J., Fr\'emat, Y., Cidale, L.} 2005, 
\textit{A\&A} 441, 235


\end{thebibliography}
\end{document}